\def \yskip{\penalty-50\vskip3pt plus 3pt minus 2pt}
\def \reference{\par \yskip \noindent \hangindent .4in \hangafter 1}
\def \abc#1#2#3#4 {\reference#1, {\sl#2}, {\bf#3}, #4}
\def \blank {\lower 5pt\hbox to 0.75in{\hrulefill}}
\def \kms{~\rm{km}~\rm{s}^{-1}}

\def \cm{~\rm{cm}}
\def \s{~\rm{s}}
\def \km{~\rm{km}}

\def \g{~\rm{g}}
\def \AU{~\rm{AU}}

\def \yr{~\rm{yr}}

\def \G{~\rm{G}}

\def \lae{\mathrel{<\kern-1.0em\lower0.9ex\hbox{$\sim$}}}
\def \gae{\mathrel{>\kern-1.0em\lower0.9ex\hbox{$\sim$}}}

\documentstyle[11pt,aasms,tighten,flushrt]{article}
\begin{document}
%\normalsize
\small

\setcounter{page}{1}
%\noindent Presented at the 180 IAU Symposium: {\it Planetary Nebulae}, 
%August 1996.
%dust1.tex
\begin{center} \bf 
MAGNETIC FIELD, DUST, AND AXISYMMETRICAL \\
MASS LOSS ON THE AGB
\end{center}
%\vspace*{2.0cm}

\begin{center}
Noam Soker\\
Department of Physics, University of Haifa at Oranim\\
%Mathematics-Physics\\
Oranim, Tivon 36006, ISRAEL \\
soker@physics.technion.ac.il 
\end{center}

%\clearpage 

\begin{center}
\bf ABSTRACT
\end{center}

I propose a mechanism for axisymmetrical mass loss on the 
asymptotic giant branch (AGB), that may account for the axially symmetric 
structure of elliptical planetary nebulae. 
 The proposed model operates for slowly rotating AGB stars, 
having angular velocities in the range of 
$10^{-4} \omega _{\rm Kep} \lae \omega \lae 10^{-2} \omega_{\rm Kep}$,
where $\omega_{\rm Kep}$ is the equatorial Keplerian angular velocity.
 Such angular velocities could be gained from a planet companion of
mass $\gae 0.1 M_{\rm Jupiter}$, which deposits its orbital angular
momentum to the envelope at late stages, or even from single stars 
which are fast rotators on the main sequence.   
The model assumes that dynamo magnetic activity  results in the 
formation of cool spots, above which dust forms much easily.  
 The enhanced magnetic activity toward the equator results in a higher
dust formation rate there, and hence higher mass loss rate. 
 As the star ascends the AGB, both the mass loss rate and magnetic activity
increase rapidly, and hence the mass loss becomes more asymmetrical,
with higher mass loss rate closer to the equatorial plane. 

\noindent 
%{\it Subject heading:}   % to APJ

%\clearpage 

% ======================================================================
\section{INTRODUCTION}
% ======================================================================

 The structures of planetary nebulae (PNs), among other arguments, suggest 
 that many AGB stars, the progenitors of PNs, have a high mass loss rate at 
the end of their AGB phase, in what is termed {\it superwind}.  % (Renzini 1981)
 The mechanism responsible for the high mass loss rate is thought
to be radiation pressure on dust (e.g., Knapp \& Morris 1985),
with some contribution from molecule acceleration 
(Johnson, Alexander, \& Bowen 1995).
 The rapid growth of the mass loss rate at the end of the AGB has been
suggested to result from the increase of the density scale height above
the photosphere (Bedijn 1988; Bowen \& Wilson 1991).
 Since the density scale height $H$ appears in the argument of an
exponential term in the expression for the density,
$\rho(r) \propto e^{-r/H}$, both the density at the location of dust 
formation and mass loss rate increase rapidly as $H$ increases.

 In many elliptical PNs the inner shell, which was formed from the superwind,
deviates more from sphericity than the outer region, which was formed
from the regular slow wind (prior to the onset of the superwind).
 In extreme cases the inner region is elliptical while the outer
region (outer shell or halo) is spherical (e.g., NGC 6826).
 In addition, most ($\sim 75 \%$) of the 18 spherical PNs 
(listed in Soker 1997 table 2) 
do not have superwind, but just an extended spherical halo.  
 The correlation between the onset of the superwind and the
increase in the asymmetry of the wind is not perfect, and in some cases 
the inner and outer region both have a similar degree of asymmetry
(e.g., NGC 7662). 
 The correlation between the superwind and highly non-spherical mass loss, 
although not perfect, is a challenge to any model for the formation of 
elliptical PNs and for the superwind's high mass loss rate.

 In an earlier paper (Soker 1995) I proposed that this correlation
results from a late interaction of the progenitor AGB star with a binary
companion, in most cases planet or brown dwarf companions (Soker 1997).
 As the star expands along the AGB the ratio of its radius $R_\ast$
to the orbital separation $a$ increases.
 Since the tidal spiraling-in time goes as $(a/R_\ast)^8$, there is a rapid 
transition from negligible tidal interaction to strong tidal interaction.
 The tidal interaction results in the spinning up of the envelope, and in
most cases the companion enters the envelope and forms a common envelope.
 The rotation of the envelope results in both axisymmetrical mass loss
and a higher mass loss rate.
 This model requires that the companion be at the right orbital separation
range: it cannot be too close in order to avoid interaction when the
star is on the red giant branch (RGB), and it cannot be too wide, 
otherwise no tidal interaction will occur. 
 Main sequence stars of mass $M_{\rm {MS}} \lae 2 M_\odot$ expand to large
radii on the RGB, and therefore are likely to interact with their companion
already on the RGB (Soker 1998).
 Therefore, the late tidal interaction model of Soker (1995) can be applied
only to stars having main sequence mass of $M_{\rm {MS}} \gae 2 M_\odot$.

 In the present paper I propose a different mechanism to explain the 
correlation of superwind and axisymmetrical mass loss.
 In this model I assume that a weak magnetic field forms cool stellar
spots on the surface of the AGB star.
 The cooler spots facilitate the formation of dust closer to the stellar
surface, enhancing the mass loss rate (Frank 1995).
{{{ Enhanced dust formation above cool regions of red giants 
was assumed by Schwarzschild (1975), where in his model the cool regions  
are formed by very large convective elements. }}}
{{{{ Frank (1995) conducted a detail study of dust formation above cool 
AGB starspots, and showed that the mass loss rate above the spots
increases, though the terminal wind velocity does not change much. 
Frank (1995), though, does not discuss the source of the cool starspots. }}}}
 A slow rotation, which is required for the amplification of the magnetic
field, forms a dipolar field which has a stronger magnetic activity
near the equatorial plane; hence the axisymmetrical mass loss.
 One advantage of the proposed model is that only a slow rotation on
the AGB is required, which can be acquired from a gas giant planet, 
or even by singly evolving stars (i.e., no further spin-up during the post
main sequence evolution is required), if they are rapid rotators on the 
main sequence. 
 The model proposed in the present study is built to explain the 
axisymmetrical mass loss from the progenitors of elliptical PNs. 
The model does not apply to bipolar PNs 
(PNs that have two lobes with an `equatorial' waist between them),  
which seem to result from stellar binary systems 
(Livio, Salzman \& Shaviv 1979; Morris 1987; Corradi \& Schwarz 1995; 
Soker 1998; Mastrodemos \& Morris 1998; and more references in these papers).  
 In $\S 2$ I summarize a number of previous papers dealing with dust 
formation and magnetic activity, which are important ingredients of the model.
 In $\S 3$ I present the phenomenological model,
and in $\S 4$ I discuss the results. 
 The summary is in $\S 5$. 
 
% ======================================================================
\section{BASIC INGREDIENTS OF THE MODEL}
% ======================================================================

% ======================================================================
\subsection {Dust}
% ======================================================================

 Bedijn (1988) and Bowen \& Wilson (1991) suggest that the rapid increase
in the mass loss rate toward the end of the AGB phase results from the 
increase in the density scale height above the photosphere.
 The density scale height is given by $H \propto R_\ast^2 T/ M_{\ast}$, 
and the density profile is $\rho(r) \propto \exp (-r/H)$, 
where $R_\ast$ is the stellar radius, $M_\ast$ its mass and $T$ 
the temperature above the photosphere.
 The density at the place where the winds starts, in the case of radiation 
pressure on dust this being the dust condensation radius, 
is given by
\begin{equation}
\rho_c \simeq  \exp (-A M_\ast / R_\ast), 
\end{equation}
where $A$ is a constant $A \sim {\rm few} \times 10^3 M_\odot ^{-1} R_\odot$  
(Bedijn 1988; Bowen 1988). 
 The mass loss rate is proportional to this density. 
 As a star ascends the AGB, $R_\ast$ increases and, because of mass loss,  
$M_\ast$ decreases, and both enhance $H$ and hence the density $\rho_c$
and the mass loss rate. 
 Bowen \& Wilson (1991) argue that dust formation is not required to 
explain the rapid increase in the mass loss rate, but it is required 
to obtain high mass loss rates as observed on the upper AGB. 
 In the present study I assume that radiation pressure on dust is the
mechanism responsible for the high mass loss rate.

 Dorfi \& H\"ofner (1996) proposed a model based on dust to explain 
axisymmetrical mass loss. 
 Their model requires a large amount of angular momentum, despite what
they term ``slow rotation''. 
 Although it is quite plausible that rotation together with radiation 
pressure on dust form axisymmetrical mass loss, I show now that their 
model {\it must} incorporate a binary companion to spin-up the envelope.  
 In order for their proposed scenario to work, Dorfi \& H\"ofner (1996) 
require the angular velocity of their AGB star, of radius $R = 500 R_\odot$,
to be $\gae 10 \%$ of the Keplerian angular velocity. 
  We hardly find main sequence stars with such high rotational velocity; 
there is no way a {\it singly} evolved AGB star can obtain this rotational 
velocity.
  Approximating the density profile on the AGB by $\rho \propto r^2$,
where $r$ is the radial distance from the center of the star, 
we find the envelope's moment of inertia to be 
$I_{\rm {env}} = (2/9) M_{\rm {env}} R_\ast ^2$, 
where $M_{\rm {env}}$ is the envelope's mass. 
 With this envelope density profile  the angular momentum of the envelope 
decreases with mass loss as $J \propto M_{\rm {env}}^3$ 
(Harpaz \& Soker 1994). 
 Since most of the mass loss occurs on the AGB, we find that the
ratio of envelope angular velocity on the upper AGB to the Keplerian 
(critical) angular velocity for a single star evolution is given by 
\begin{eqnarray}
\left( {{\omega}\over{\omega_{\rm Kep}}} \right)_{\rm Single-AGB}
\simeq 
10^{-4}     
\left( {{\omega}\over{0.1 \omega_{\rm Kep}}} \right)_{\rm MS}
\left( {{R_{\rm MS}}}\over{0.01R_{\rm {AGB}}} \right)^{1/2} 
\left({{M_{\rm {env}}} \over {0.1 M_{\rm {env0}}}} \right)^ {2},
\end{eqnarray}
where the subscript MS means that the quantity is taken
at the end of the main sequence, and $M_{\rm {env0}}$ is the
envelope mass at the beginning of the AGB. 
The contraction of the star from the AGB during the post-AGB phase
will result in a further spin down (Soker 1992). 
 This is because the high mass loss rate causes mass shells to move
outward in the convective envelope, despite the contraction of the
star.  
 Much more massive stars which retain massive envelopes may spun-up 
when they move to the left of the HR diagram during their
blue loop (Heger \& Langer 1998).
This is caused by the transfer of angular momentum from the inner to the 
outer convective region, as the inner radiative region of the envelope
expands (Heger \& Langer 1998). 
 Since the contraction of post AGB stars is due to the depletion of 
the very low-mass post AGB envelope by the wind, which carries most of
the envelope's angular momentum, 
this mechanism cannot work for contraction beyond the AGB 
(Soker 1992).

  The orbital separation of a low mass secondary when tidal interaction 
becomes significant is $a \sim 5 R_\ast$ (Soker 1998).
 If the secondary deposits all its orbital angular momentum to the
envelope of mass $M_{\rm env}$, the ratio of envelope angular velocity 
$\omega$ to the surface Keplerian angular velocity $\omega_{\rm Kep}$ 
is given by 
\begin{eqnarray}
{{\omega}\over{\omega_{\rm Kep}}} \simeq 
0.1   
\left( {{M_2}\over{0.01M_{\rm env}}} \right) 
\left({{a} \over {5R_\ast}} \right)^ {1/2},
\end{eqnarray}
assuming that the entire envelope rotates uniformly. 
We conclude that in order to spin-up the envelope as required by 
Dorfi \& H\"ofner the secondary mass should be 
$M_2 > 0.01 M_{\rm {env}}$. 
 However, as Harpaz \& Soker (1994) show, the envelope's specific angular
momentum of an AGB star decreases with mass loss as 
$L_{\rm env}/M_{\rm env} \propto M_{\rm env}^{2}$. 
 Therefore, to supply the angular momentum for a longer time, 
the companion mass should be much larger than $0.01 M_\odot$, i.e.,
a brown dwarf or a low main sequence star. 
  Other effects that such a companion can cause (Soker 1997) should
then be considered as well. 
  In the present paper the required angular velocity is much smaller,
since I incorporate magnetic fields, and a planet companion of mass 
$\sim 0.1 M_{\rm Jupiter}$ is sufficient to spin-up the envelope. 

 We now show the problems with direct effects of rotation,
e.g., centrifugal force, for more general cases. 
 The centrifugal force on the equatorial plane increases the scale height 
by a factor of $[1-(\omega/\omega_{\rm {Kep}})^2]^{-1}$. 
 The scale height along the polar directions does not change.  
 Inserting this factor in equation (1) for the equatorial plane we get  
$(\rho_c)_e \simeq  \exp \left[ -A M_\ast R_\ast^{-1} 
(1-[\omega/\omega_{\rm {Kep}}]^2) \right]$.
The ratio of equatorial to polar density is then
\begin{equation}
{{(\rho_c)_e}\over{(\rho_c)_p}} \simeq 
\exp [A M_\ast R_\ast^{-1} (\omega/\omega_{\rm {Kep}})^2].
\end{equation}
The strong dependence of the density contrast, and hence mass loss rate
contrast, between the equatorial and the polar directions on the angular 
velocity is what Dorfi \& H\"ofner (1996) find in their numerical calculations.
 Numerically, the results of Dorfi \& H\"ofner (1996),
for $M_\ast = 1M_\odot$ and $R_\ast = 500 R_\odot$, fit a value of 
$A \simeq  2 \times 10^4 M_\odot ^{-1} R_\odot$ rather
than $A = 5 \times 10^3 M_\odot ^{-1} R_\odot$ found by Bedijn (1988). 
 Two problems for a model based on the centrifugal force emerge from 
equation (4). 
 First, as the star ascends the AGB and loses mass, $M_\ast$ decreases,
$R_\ast$ increases, and because of expansion and mass loss
$(\omega/\omega_{\rm {Kep}})^2$ decreases.
 Therefore, from equation (4), the density contrast between the equatorial
and polar directions {\it decreases}, contrary to what is observed
in most PNs. 
 The second problem is the required angular velocity. 
 Inserting  $A = 2 \times 10^4 M_\odot ^{-1} R_\odot$ in equation (4), 
we find the angular velocity required to obtain a density contrast 
$q \equiv (\rho_c)_e/(\rho_c)_p$ to be 
\begin{equation}
{{\omega}\over{\omega_{\rm {Kep}}}} = 0.08
\left( {{\ln q}\over{\ln 2}} \right)^{1/2}
\left( {{M_\ast}\over{M_\odot}} \right)^{-1/2} 
\left( {{R_\ast}\over{200 R_\odot}} \right)^{1/2} .
\end{equation}
As mentioned above, such a high angular velocity requires a
companion to spin-up the envelope, at least a brown dwarf, and probably
a low mass stellar companion. 
% Taking into account the mass loss after the angular momentum deposition, 
% I find that planets do not have enough angular momentum to keep AGB 
% envelopes to rotate at such high angular velocities.  
 
% ======================================================================
\subsection {Magnetic Field}
% ======================================================================
 Direct magnetic effects, through magnetic tension and/or pressure,
have been suggested to determine the mass loss geometry from AGB stars 
(Pascoli 1997 and references therein), or to influence the circumstellar
structure during the PN phase (Chevalier \& Luo 1994; Garcia-Segura 1997).
 As I show in this subsection, these models {\it must} incorporate
a binary companion to substantially spin-up the envelope. 
 These models will be compared later to the non-direct magnetic
effects proposed in the present study. 
 
 In several papers, Pascoli (1997, and references therein) has suggested
that magnetic activity plays a major role in determining the mass loss 
rate and geometry from AGB stars.
 Although Pascoli's general idea of magnetic activity might be right,  
there are a number of specific problems with the details presented 
in his models (Soker \& Harpaz 1992). 
 Pascoli (1997), for example, proposes that the magnetic field formation,
through dynamo mechanism, occurs close to the core of AGB stars. 
 He then assumes that the core surface, of radius $R_c = 10^9 \cm$, 
rotates at a velocity of $v_c = 100 \kms$, and that the envelope 
convection penetrates down to the core's surface. 
 Both assumptions seem to me problematic. 
 First, detailed models of AGB stars show that there is a large 
radiative zone above the core (e.g., Soker 1992), and that the envelope's 
convective region starts only at $r \sim 1 R_\odot$. 
 Second, the convection acts to enforce a constant angular velocity
$\omega (r) \simeq \omega_0$ throughout the convective region,
contrary to Pascoli's assumption of constant specific angular 
momentum $\omega(r) \propto r^{-2}$. 
 Therefore, even if the convection penetrates down to the core's surface, 
it will substantially reduce the angular velocity compared with the value 
assumed by Pascoli. 
  Therefore, the rotation velocity assumed by Pascoli requires that
the AGB star be spun-up by a companion.
 Further indication for the too high velocity assumed is the finding that
single white dwarfs rotate very slowly, $v_{\rm {rot}} \ll  50 \km \s^{-1}$
(Heber, Napiwotzki, \& Reid 1997).
Using lower rotation velocities, as indicated above, will result in 
a much weaker magnetic activity than that assumed by Pascoli. 
  
 The second model I find to require substantial envelope spin-up is that of 
Garcia-Segura (1997; see also Garcia-Segura {\it et al.} 1998), which is 
an extension of the model proposed by Chevalier \& Luo (1994; 
see also Chevalier 1995). 
 This model is based on the tension of the toroidal component of the 
magnetic field in the wind: the wind in the transition from the AGB to the
PN phase or the fast wind during the PN phase. 
 Close to the star the magnetic pressure and tension are negligible compared
with the ram pressure and thermal pressure of the wind. 
 As the wind hits the outer PN shell, which is the remnant of the slow wind,
it goes through a shock and slows down, and the toroidal component 
of the magnetic field increases substantially. 
 This may result in the magnetic tension and pressure becoming the 
dominant forces near the equatorial plane.   
 In particular, the magnetic tension pulls toward the center and reduces 
the effective pressure in the equatorial plane. 
 According to this model (Chevalier \& Luo 1994; Garcia-Segura 1997), 
then, the equatorial plane will be narrow, leading to an elliptical 
or bipolar PN. 

 The efficiency of this model is determined by the parameter 
(Chevalier \& Luo 1994) 
\begin{equation}
\sigma
=
\left( {{B_s^2r_s^2}\over{\dot M_w v_w}} \right)
\left( {{v_{\rm {rot}}}\over{v_w}} \right)^2 
= {{\dot E_B}\over{\dot E_k}} 
\left({{v_{\rm {rot}}}\over{v_w}} \right)^2, 
\end{equation}
where $B_s$ is magnetic field intensity on the stellar surface,
$r_s$ the stellar radius,  $\dot M_w$ the mass loss rate into the wind, 
$v_w$ the terminal wind velocity, and $v_{\rm {rot}}$ the equatorial 
rotational velocity on the stellar surface.
  In obtaining the second equality the expressions for the magnetic energy 
luminosity  $\dot E_B=4 \pi r_s^2 v_w B_s^2 / 8 \pi$ and for the kinetic 
energy luminosity $\dot E_k = \dot M_w v_w^2/2$ were used. 
 For the model to be effective it is required that $\sigma \gae 10^{-4}$,
but a typical value of $\sigma \simeq 0.01$ is used by 
Garcia-Segura (1997). 
 For the sun $\sigma \simeq 0.01$ and 
$(v_{\rm {rot}}/v_w)^2 \simeq 2 \times 10^{-5}$ (Chevalier \& Luo 1994). 
 However, in the sun it is magnetic activity which determines the mass loss
rate, as we see from the ratio $\dot E_B/\dot E_k \simeq 500$.  
 It is commonly assumed that radiation pressure drives the winds of
central stars of PNs, and that pulsation together with radiation pressure
drives the wind of AGB stars, and red giants in general. 
 Therefore, the sun is not a good example of this model for singly evolved 
stars. 
 If magnetic energy does not drive the wind then 
$\dot E_B/\dot E_k \lae 1$, and the model of magnetic shaping requires 
$ v_{\rm {rot}}/v_w \gae 0.01$.  
 Such rotation velocity is {\it impossible} for singly evolved AGB or
post-AGB stars to attain (Harpaz \& Soker 1994; $\S 2.1$ above). 
 
 It is clear that in order for the magnetic shaping model of
Chevalier \& Luo (1994) and Garcia-Segura (1997) to be of any significance, 
a {\it substantial} spinning-up by a binary companion is required. 
 But even if this condition is met, I find two other problems with this 
model concerning shaping on a large scale.
 The two problems result from MHD instabilities. 
 Other MHD instabilities might exist as well (Livio 1995). 
(a) The magnetic shaping model requires that the magnetic field lines
circle the central star in the equatorial plane. 
 However, it is not clear that this will be the case. 
 Because of MHD instability the magnetic field escapes from the sun 
in non-axisymmetrical magnetic flux loops (e.g., Bieber \& Rust 1995; 
Caligari, Moreno-Insertis, \& Sch\"ussler 1995). 
 Many flux loops which escape the sun  do not circle the sun. 
 It is possible, though, that when the magnetic pressure is low  
more flux loops will circle the star. 
(b) For the magnetic field to be of any significance, it should be 
regenerated by a stellar dynamo. 
 The idealized toroidal magnetic field that results from a dynamo has 
opposite directions in the two stellar hemispheres (e.g., Bieber \& Rust 1995).
 Therefore, as the magnetic pressure becomes dominant after the wind
slows down, I expect that reconnection of magnetic field lines close 
to the equatorial plane will occur. 
 This will reduce the magnetic pressure and tension near the equatorial 
plane, and as a consequence will reduce the efficiency of the model. 
 
 The last two problems relate only to the large-scale shaping.
 Magnetic field may still be strong but with a short coherence length. 
  I think that when substantial spinning occurs and if dynamo activity 
becomes efficient, magnetic fields may play a substantial role in 
small-scale shaping, e.g., MHD instability modes on small scales,
fragmantation of regions in which SiO masers exist (Hartquist \& Dyson 1997),  
and non-thermal radio emission (Dgani \& Soker 1998).

% ======================================================================
\section {THE MODEL}
% ======================================================================
% ======================================================================
\subsection {Outline of the Model}
% ======================================================================

 In order to circumvent the fast rotation required by the models described
in the previous subsections, but still using the popular model of radiation
pressure on dust particles for the superwind, I propose the following
scenario. 
 I assume a weak magnetic activity, i.e., the magnetic tension or pressure
never become the dominant force globally, neither on the surface 
nor at large distances from the star.
 As in the sun, the magnetic activity results in the formation of 
{\it cool spots}. 
 Above these spots, I suggest, dust forms much more easily, i.e., faster and
closer to the stellar surface (Frank 1995). 
 The formation of dust closer to the stellar surface means higher density
there, and hence higher mass loss rate, as summarized in $\S 2.1$
and as was shown by Frank (1995). 
 A rotation is required in this scenario to amplify the magnetic field,
but a much slower rotation than what is required in the models mentioned 
earlier. 
 It might even be the case that fast rotating singly evolved
stars can possess such rotation velocities 
(i.e., no further spin-up is required, $\S 3.2$). 
{{{  Fastly rotating main sequence stars have mass of 
$M_{\rm MS} \gae 2 M_\odot$.
This is the mass range of progenitors of bipolar PNs (e.g., 
Corradi \& Schwarz 1995), and of stars that form type I PNs
(Torres-Peimbert \& Peimbert 1997).  
 However, not all massive stars form bipolar PNs, and not all
type I PNs are bipolars  (Corradi \& Schwarz 1995).
 Soker (1998) estimates that only $\sim 40 \%$ of stars having 
main sequence mass of $2 M_\odot  \lae M_{\rm MS} \lae 8 M_\odot$ 
form bipolar PNs. 
 If bipolar PNs are formed from interaction with stellar companions
(Morris 1987; Livio  1997; Soker 1998), then most of the rest 
massive stars will form elliptical PNs. 
 The massive progenitors of elliptical PNs may rotate fast when 
arriving to the AGB through interaction with planets (Soker 1996), 
or they may retain enough angular 
momentum from their main sequence phase. }}}

The dynamo amplification results in a higher activity closer to the 
equatorial plane, therefore a higher mass loss rate there. 
 The increase in the mass loss rate contrast between the polar and 
equatorial directions as the star ascends the AGB is explained 
by the rapid decrease in the thermal pressure of the atmosphere,
and the formation of dust at higher densities ($\S 2.1$).  
 The cool spots mechanism means that the enhanced mass loss rate in the 
equatorial plane will be sporadic, though over a long time period 
it will be averaged to form a smooth nebula.  
 Episodic, local formation of dust occurs close to the surface of
R Coronae Borealis (RCB) stars (Clayton 1995 and references therein). 
 RCBs are hydrogen-deficient carbon-rich supergiants, thought
to result from the final helium shell flash (Renzini 1990).
 The common interpretation of observations is that dust forms 
close to the stellar surface ($< 2 R_\ast$), on an area covering
only a fraction of the surface, and that it occurs at irregular intervals.
 RCBs are not rapid rotators and there is no evidence for binarity
(Clayton 1995). 
 Whitelock {\it et al} (1997) find erratic behavior similar to that of 
RCB stars in three Carbon-rich variable AGB stars. 
 The ejection of carbon-rich dust occurs in a preferred direction,
which they speculate is the equatorial plane. 
{{{{ Frank (1995), whose model is built for carbon rich stars,
brings further evidences to support clumpy mass loss from AGB stars. }}}}
 According to the model proposed here, magnetic activity, which requires 
only slow rotation, forms cool spots on which dust is formed. 
{{{ The formation of cool spots by magnetic activity is
seen on the surface of the sun. 
 The assumption of enhanced dust formation above cool regions of 
red giants was used by Schwarzschild (1975). 
In Schwarzschild's model the cool regions are formed by very 
large convective elements. }}}
 Another hydrogen deficient object is the PN A30 {{{ (Jacoby \& Ford 1983). }}} 
 This PN has a large, almost spherical, halo, with optically bright, 
hydrogen-deficient, blobs in the inner region. 
 The blobs are thought to result from a late helium shell flash.
%(REFERENCE--from Chu??)
 These blobs are arranged in a more or less axisymmetrical shape.  
It is possible that during the formation of the halo dust was forming
far from the stellar surface.  
 {If} after the helium flash the formation of dust occurred closer to the 
stellar surface, the process became more vulnerable to magnetic activity, 
resulting in the axisymmetrical mass loss. 

 A hint of the presence of magnetic fields in the atmospheres of some
AGB stars comes from the detection of X-ray emission from a few
M giants (H\"unsch {\it et al.} 1998). 
 An even stronger motivation for the proposed model is the observation
of a magnetic field in the extended atmosphere of the Mira variable
TX Cam (Kemball \& Diamond 1997). 
 Kemball \& Diamond (1997) find the intensity of the magnetic 
field in the locations of SiO maser emission to be $B \lae 5 G$. 
 These local emission regions form a ring around TX Cam at a 
radius of $4.8 \AU \simeq 2 R_\ast$.
  Kemball \& Diamond (1997) mention the possibility that the mass 
loss occurs in a preferred plane.
 They also suggest that ``The fine-scale features [of the Maser 
image] are consistent with local outflows, flares or prominences, 
perhaps coincident with regions in which localized mass loss has 
taken place.''

% ======================================================================
\subsection{The Magnetic Activity}
% ======================================================================
 Because of the lack of basic theories for (a) dynamo operation
in AGB stars, where the convective overturn time is shorter than 
the rotation period, (b) formation of stellar cool spots through
the action of magnetic flux tubes, and (c) dust formation close
to the stellar surface, the required magnetic activity will be 
studied phenomenologically. 
 For this I take the following:
\newline
(1) From the sun it is known that the magnetic fields in sunspots 
($B_{\rm spot} \sim 2000 \G$) are $\sim 10^3$ times stronger than the average
magnetic field ($B_{\rm av} \sim 2 \G$). 
 It is possible that even if the solar average photospheric
magnetic field had been much lower,
the magnetic field pressure inside a sunspot could still have reached the 
photospheric pressure, i.e., the ratio would have been much larger 
than $\sim 10^3$.
 {{{Since convection seems to have a major role in concentrating the
magnetic field to form sunspots (Priest 1987, chap. 8),
and AGB stars have strong convection, I assume the ratio of
spots to average magnetic field to be larger than that in the sun by
an order of magnitude. 
  This assumption will have to be tested when more detailed dynamo 
models become available.                     }}
 The maximum value of magnetic pressure that can
be attained inside stellar spots is therefore
\begin{equation}
B_{\rm spot} = \eta B_{\rm av},  
\end{equation}
with $\eta \sim 10^4$. 
\newline
(2) For a spot to be significantly cooler than its surroundings I assume 
that the magnetic pressure is of the order of the photospheric pressure
\begin{equation}
{{B_{\rm spot}^2}\over{8 \pi}} \simeq P_{\rm ph}.
\end{equation}
\newline
(3) The pressure of the photosphere is taken to be (e.g., 
Kippenhahn \& Weigert 1990, $\S 10.2$)
\begin{equation}
P_{\rm ph} = {{2}\over{3}} {{G M}\over{R_\ast^2}}
{{1}\over{\kappa}},
\end{equation}
where $\kappa \simeq 3 \times 10^{-3} \cm^2 \g^{-1}$ is the opacity
at the surface of the AGB star. 
This matches the photospheric pressure of numerical models 
quite well (e.g., Soker 1992).

 Using the three equations (7)-(9), which serve as the basic assumptions
of the model, gives for the average magnetic intensity 
required to form AGB stellar spots
\begin{equation}
B_{\rm av} \gae 
4 \times 10^{-3}        
\left( {{M}\over{1M_\odot}} \right)^{1/2} 
\left( {{R_\ast}\over{300 R_\odot}} \right)^{-1}
\left( {{\eta}\over{10^4}} \right)^{-1} 
\kappa_3^{-1/2} 
\G,
\end{equation}
where $\kappa_3 \equiv \kappa/3 \times 10^{-3} \cm^2 \g^{-1}$.
  Taking $B_{\rm av}$ from equation (10) in the expressions for the 
magnetic energy luminosity 
$\dot E_B=4 \pi R_\ast^2 v_w B_{\rm av}^2 / 8 \pi$, and using  the kinetic 
energy luminosity $\dot E_k = \dot M_w v_w^2/2$, give the magnetic 
activity required for the formation of AGB stellar spots
\begin{equation}
\left( {{\dot E_B}\over{\dot E_k}} \right)_{\rm spot}
\gae   10^{-4} 
\left( {{M}\over{1M_\odot}} \right) 
\left( {{\dot M_w}\over{10^{-6} M_\odot \yr ^{-1}}} \right) ^{-1}
\left( {{v_w}\over{10 \km \s^{-1}}} \right) ^{-1}
\left( {{\eta}\over{10^4}} \right)^{-2} 
\kappa_3^{-1}. 
\end{equation}

 As mentioned earlier, there is no basic theory to predict the
magnetic activity of stars. 
 Soker \& Harpaz (1992) estimated, through a phenomenological 
study, the magnetic activity of rotating AGB stars. 
 They assume that the amplification time of the magnetic field $\tau_a$
is equal to the rotation period. 
 Using their equations (2.7) and (2.8), with the assumption of amplification
time equal to the rotational period, with minor modifications
of scalings, the ratio of magnetic to wind energy loss is 
\begin{equation}
\left( {{\dot E_B}\over{\dot E_k}} \right)_{\rm dynamo}
\simeq 10^{-4}
\left( {{\omega}\over{10^{-4} \omega _{\rm Kep}}} \right) ^{2}
\left( {{R_\ast}\over{300 M_\odot}} \right) ^{-2}
\left( {{M}\over{1M_\odot}} \right) 
\left( {{M_{\rm env}}\over{0.1M_\odot}} \right) 
\left( {{\dot M_w}\over{10^{-6} M_\odot \yr ^{-1}}} \right) ^{-1}
\left( {{v_w}\over{10 \km \s^{-1}}} \right)^{-2} .
\end{equation}
 By comparing equation (11) with equation (12), it turns that according 
to the dynamo activity estimated by Soker \& Harpaz (1992), the required 
AGB stellar angular velocity is $\sim 10^{-4}$ times the Keplerian surface 
velocity. 
 The dynamo activity derived by Soker \& Harpaz (1992), however, follows 
ideas developed for the solar dynamo, where the rotation velocity is 
faster than the convective velocity, opposite to rotating AGB stars.
 It is therefore possible that the amplification in the strong convective 
envelope of AGB stars is more efficient.  
 I propose that the model presented here is effective for 
\begin{equation}
\left( {{\omega}\over{\omega _{\rm Kep}}} \right)_{\rm spot-model}
\gae 10^{-4} .
\end{equation}

% ======================================================================
\section{DISCUSSION}
% ======================================================================
 
Several points emerge from the analysis presented in the previous section.
\newline
(1) By comparing equation (13), for the required angular velocity,
with equation (2), for the angular velocity of singly evolved AGB stars,
it turns that single stars may possess high enough angular momentum to
account for axisymmetrical mass loss, according to the proposed model. 
 However, this requires that the surface angular velocity on the main 
sequence be $\gae 5\%$ of the Keplerian angular velocity. 
 The sun, for example, is on the border;
it rotates very slowly, but the fraction of the envelope that will be 
retained on the upper AGB, when the envelope mass is, say, 
$0.05 M_\odot$, is larger than the scaling used in equation (2). 
 Other models, such as those described in $\S 2$, require substantial
spin-up by a companion.  
 This led me (Soker 1996; 1997; and references therein) % ApJ Supp
to propose that planets are commonly present around stars.  
 However, if in the next few years the results of the intensive 
planet search projects is that $\ll 50 \%$ of all stars have 
planets, then there will be a need for a model of efficient axisymmetrical 
mass loss for singly evolved stars. 
 In addition, the mechanism should account for the increase in the
degree of asymmetry toward the termination of the AGB evolution (see $\S 1$).
The mechanism proposed in the present study may be such a mechanism,
as is the mechanism of mode-switch to nonradial oscillations,  
proposed by Soker \& Harpaz (1992). 
 It is hard to tell from this preliminary study if the mechanism 
proposed here can be applied to a large fraction of singly evolved stars,
since most main sequence stars may not rotate fast enough.
However, the mechanism is quite effective if a planet companion of mass 
$\gae 0.1 M_{\rm Jupiter}$ enters the envelope at late stages of the 
AGB or the RGB and spins-up the envelope. 

2) From equation (2) we see that the angular velocity decreases rapidly
as the envelope mass decreases toward the termination of the AGB.
In the present model this decrease is more than compensated for by
the increase of the vulnerability of dust formation and photospheric
conditions to the magnetic activity (point 6 below). 

3) From equation (6) and the requirement $\sigma \gae 10^{-4}$, we find 
for the magnetic model of Chevalier \& Luo (1994) and Garcia-Segura (1997), 
$ \dot E_B/\dot E_k \gae  (v_{\rm {rot}}/0.01v_w )^{-2}$. 
 It should be noted that equation (11) refers to AGB stars, while 
the model of CL refers to post-AGB stars, which are much
hotter and smaller, but are left with a small envelope mass, and are
expected to rotate very slowly according to equation (2). 
 It is clear from comparing equation (11) with the value given above 
that the presently proposed model is much less demanding on the 
magnetic activity and/or rotation velocity on the AGB and beyond,
compared with the models described in $\S 2$. 
 
4) The required magnetic activity $\dot E_B$ (eq. 10) does not depend on
the mass loss rate.

5) Because the magnetic energy released through the photosphere
is much below both the kinetic and thermal energy carried by the wind, 
the magnetic activity will not heat the region above the photosphere,
except perhaps in localized regions where the magnetic energy
becomes extremely strong. 
 This is the opposite situation to that in the solar corona, where the
magnetic energy is larger than the wind's kinetic energy,
$\dot E_B/\dot E_k \simeq 500$.  

6) How is the increase in the asymmetry degree with the increase of 
the mass loss rate toward the end of the AGB explained in the 
present model?  
 The answer suffers from the lack of a theory to 
predict the formation of cool spots and the formation of dust above 
these cool spots. 
 But let me try to propose two plausible explanations. 
\newline 
6.1) As the star ascends the AGB it expands and loses mass. 
 Both processes cause a rapid decrease of the density and pressure 
in the envelope below the photosphere (e.g., Soker 1992). 
 This is the region where the magnetic field behavior determines the
appearance of the magnetic field in the photosphere. 
 Therefore, the low density and pressure may result in a much more
violent magnetic behavior, leading to the appearance of larger and
more frequent stellar cool spots. 
 A very low envelope density and pressure exist in extended envelopes 
formed by late final helium shell flash, such as proposed for 
R Coronae Borealis stars and the PN A 30 ($\S 3.1$).  
 In these types of objects dust formation may be facilitated by the
high metalicity as well. 
\newline 
6.2) For the mechanism proposed here to be effective, dust formation 
should occur close to the spot, since the temperature difference between
the spot and its surroundings is smoothed and vanishes as the distance 
from the stellar surface increases (Frank 1995). 
 As the star gets cooler, as is the case for stars ascending the AGB,
the dust forms closer to the star, and hence the cool spots influence
dust formation more strongly. 
 The transition to a carbon-rich AGB star on the upper AGB 
may also make dust formation easier (Frank 1995).  

7) The higher mass loss rate during the superwind phase is explained both 
by the increase of the density scale height 
(Bedijn 1988; Bowen \& Wilson 1991; $\S 2.1$), which influences the 
global mass loss rate, and by the increase of the magnetic activity, 
with increasing activity toward the equatorial plane. 
 Therefore, both the axisymmetrical mass loss, which results from the 
magnetic activity, and the total mass loss rate increase as the star
evolves along the AGB. 
  The existence of spherical PNs which have no superwind, 
$\sim 75 \%$ of all spherical PNs, has two plausible explanations 
in the frame of the proposed model.
 The first is that the progenitors of these PNs lost their entire envelope
during the early AGB.  
 The second is that the magnetic activity is responsible for the 
superwind in some cases, and the progenitors of the spherical PNs
were rotating too slowly to support magnetic activity. 

8) The solar magnetic activity has a cycle of 11 years period.  What
will be the consequences, in the proposed model, if rotating AGB 
stars have magnetic activity cycle?
It is impossible to predict the period of such a cycle because of 
the lack of a detailed dynamo model. 
 But if exist, it will, according to the proposed model, cause oscillations
in the mass loss rate.
 I would like to speculate that the arcs, or shells, found recently 
in several PNs (e.g., CRL 2688 [Egg Nebula], Sahai {\it et al.} 1998a,b;
IRAS 17150-3224, Kwok, Su, \& Hrivnak 1998)
may result from such magnetic activity cycle. 
 In IRAS 17150-3224 the arcs are concentric, nearly circular
(Kwok {\it et al.} 1998), while in CRL 2688 
the arcs show departure from circularity, and a large fraction of them
span only small angles (Sahai {\it et al.} 1998b). 
They arcs represent almost periodic enhancement in mass loss rate,
(Kwok {\it et al.} 1998), by a factor of at least 2 
(Sahai {\it et al.} 1998a). 
 The periods are a few hundred years. 
 A magnetic activity cycle, similar to the solar cycle, but of a few
hundred years period, may explain the almost periodic, and somewhat
sporadic in direction, mass loss that formed the shells. 
 A possible problem to the proposed model is that the shells are almost
spherical, though some differences between the equatorial and polar 
direction exist. 
 Other models for the formation of the arcs 
(e.g., Harpaz, Rappaport \& Soker 1997) are discussed by Sahai {\it et al.}
(1998b).
 
% ======================================================================
\section{SUMMARY}
% ======================================================================

The main goal of the paper was to propose a mechanism for axisymmetrical 
mass loss on the AGB that: (a) increases as the star evolves along the AGB,
and (b) operates for slowly rotating AGB stars (having angular velocity 
in the range of
$10^{-4} \omega _{\rm Kep} \lae \omega \lae 10^{-2} \omega_{\rm Kep}$,
where $\omega_{\rm Kep}$ is the equatorial Keplerian angular velocity).
 Such angular velocities could be gained from a planet companion of
mass $\gae 0.1 M_{\rm Jupiter}$, which deposits its orbital angular
momentum to the envelope during the AGB phase or even much earlier
during the RGB phase, or even from single stars which are fast 
rotators on the main sequence.   
 The model is built to explain elliptical PNs, but not the more
extremely asymmetrical bipolar PNs, which are thought to be formed from 
stellar binary systems. 
 
 The proposed model assumes that dynamo magnetic activity results in the 
formation of cool spots, above which dust forms much easily.  
 The enhanced magnetic activity toward the equator results in a higher
dust formation rate there, hence higher mass loss rate. 
 As the star ascends the AGB, both the mass loss rate and magnetic activity
increase rapidly. 

 Future observations should search for magnetic fields around AGB stars,
e.g., in masers, and for sporadic mass loss episodes, such as in 
R Coronae Borealis stars.  
 An extremely important issue, which is relevant to all models of
axisymmetrical mass loss, is the detection of rotations in AGB stars. 
 On the theoretical side, it will be necessary to extend the work of
Frank (1995), i.e., calculate the density and temperature profile above 
AGB stellar spots, and from that the dust formation rate. 
{{{ Cool spots formed by large convective elements (Schwarzschild 1975) will
also have to be considered. }}}
 These are difficult tasks because of the strong pulsations in AGB stars. 
 An even more difficult task is to build a dynamo theory for slowly 
rotating AGB stars. 

{\bf ACKNOWLEDGMENTS:} 
{{{ I thank an anonymous referee for detailed comments which 
improved the presentation of the proposed model.}}}
 This research was supported in part by a grant from the University
of Haifa and a grant from the Israel Science Foundation.

\end{document}